\documentclass[conference]{IEEEtran}

\usepackage{cite}
\usepackage{graphicx}
\usepackage{epstopdf} 
\usepackage{subcaption}
\usepackage{multirow}
\usepackage{pbox}
\usepackage{amsmath}
\usepackage{amssymb}
\graphicspath{{./figures/}}
\DeclareGraphicsExtensions{.pdf,.png,.jpg,.eps}
\usepackage{xcolor}
\usepackage[official]{eurosym}
\usepackage[nolist]{acronym}
\usepackage[symbol]{footmisc}

\makeatletter
\def\footnoterule{\kern-3\p@
  \hrule \@width 2in \kern 2.6\p@} 
\makeatother

\begin{document}
\title{Spectrum Monitoring for Radar Bands using  Deep Convolutional Neural Networks\vspace{-0.2in}} 

\author{\IEEEauthorblockN{Ahmed Selim\textsuperscript{*},
Francisco Paisana\textsuperscript{*},
Jerome A. Arokkiam,
Yi Zhang,
Linda Doyle, 
Luiz A. DaSilva}  
\IEEEauthorblockN{CONNECT center, Trinity College Dublin, Ireland}
\IEEEauthorblockN{email:
\{selimam, paisanaf, a.jerome, zhangy8, linda.doyle, dasilval\}@tcd.ie}
\vspace{-0.3in}
}

\maketitle
\begin{abstract}
In this paper, we present a spectrum monitoring framework for the detection of radar signals in spectrum sharing scenarios. The core of our framework is a deep convolutional neural network (CNN) model that enables \acp{mcd} to identify the presence of radar signals in the radio spectrum, even when these signals are overlapped with other sources of interference, such as commercial \ac{lte} and \ac{wlan}. We collected a large dataset of RF measurements, which include the transmissions of multiple radar pulse waveforms, downlink \ac{lte}, \ac{wlan}, and thermal noise. We propose a pre-processing data representation that leverages the amplitude and phase shifts of the collected samples. This representation allows our \ac{cnn} model to achieve a classification accuracy of $99.6\%$ on our testing dataset. The trained \ac{cnn} model is then tested under various SNR values, outperforming other models, such as spectrogram-based \ac{cnn} models.
\end{abstract}

\footnotetext[1]{Denotes equal contribution}

\begin{acronym}
%
%
%
%
%
\acro{cbrs}[CBRS]{Citizens Broadband Radio Service}
\acro{cnn}[CNN]{Convolutional Neural Network}
\acro{dfs}[DFS]{Dynamic Frequency Selection}
\acro{esc}[ESC]{Environment Sensing Capability}
\acro{fcc}[FCC]{Federal Communications Commission}
\acro{gldb}[GLDB]{Geo-location Database}
\acro{ipm}[IPM]{Intra-Pulse Modulation}
\acro{lbt}[LBT]{Listen-Before-Talk}
\acro{lfm}[LFM]{Linear Frequency Modulation}
\acro{lte}[LTE]{Long-Term Evolution}
\acro{mcd}[MCD]{Measurement Capable Device}
\acro{mcs}[MCS]{Modulation Coding Scheme}
\acro{ntia}[NTIA]{National Telecommunications and Information Administration}
\acro{pc}[PC]{Pulse Carrier}
\acro{pf}[PF]{Pulse Frequency}
\acro{pm}[PM]{Phase Modulation}
\acro{pri}[PRI]{Pulse Repetition Interval}
\acro{pw}[PW]{Pulse Width}
\acro{rat}[RAT]{Radio Access Technology}
\acro{rats}[RATs]{Radio Access Technologies}
\acro{rem}[REM]{Radio Environment Map}
\acro{sas}[SAS]{Spectrum Access System}
\acro{su}[SU]{Secondary User}
\acro{toa}[TOA]{Time of Arrival}
\acro{tvws}[TVWS]{TV White Space}
\acro{wlan}[WLAN]{Wireless Local Area Network}
\end{acronym}

\section{Introduction}
\label{sec:introduction}

Spectrum sharing frameworks reliant on \acp{gldb} tend to leave spectrum underutilized. The reasons behind this are well understood and pointed out in the \ac{tvws} and \ac{cbrs} literature \cite{gavrilovska2014enabling,FCC2015}. \acp{gldb}, in their current form, cannot cope with the mobility of incumbents, and rely on theoretical propagation models that do not efficiently account for factors like terrain elevation and other obstructions\cite{PaisanaSurvey2014}.



Spectrum monitoring networks combined with \acp{rem} may prove to be an important tool for addressing the aforementioned issues\cite{gavrilovska2014enabling}. As \acp{rem} rely on real-time RF measurements, they can provide a more exact picture of the current spectrum occupancy across time, frequency and space.  \ac{esc}, currently under investigation for the \ac{cbrs} band, is an example of the spectral efficiency gains such an approach can provide \cite{Drocella2015}. \acp{rem} may also allow operators to optimize their resource allocations and identify potential sources of interference, and provide regulators with a tool for spectrum enforcement, without the costs of sending specialized staff to the field for interference analysis \cite{Nika2014}.

\begin{figure}
\centering
  \includegraphics[width=0.9\columnwidth] {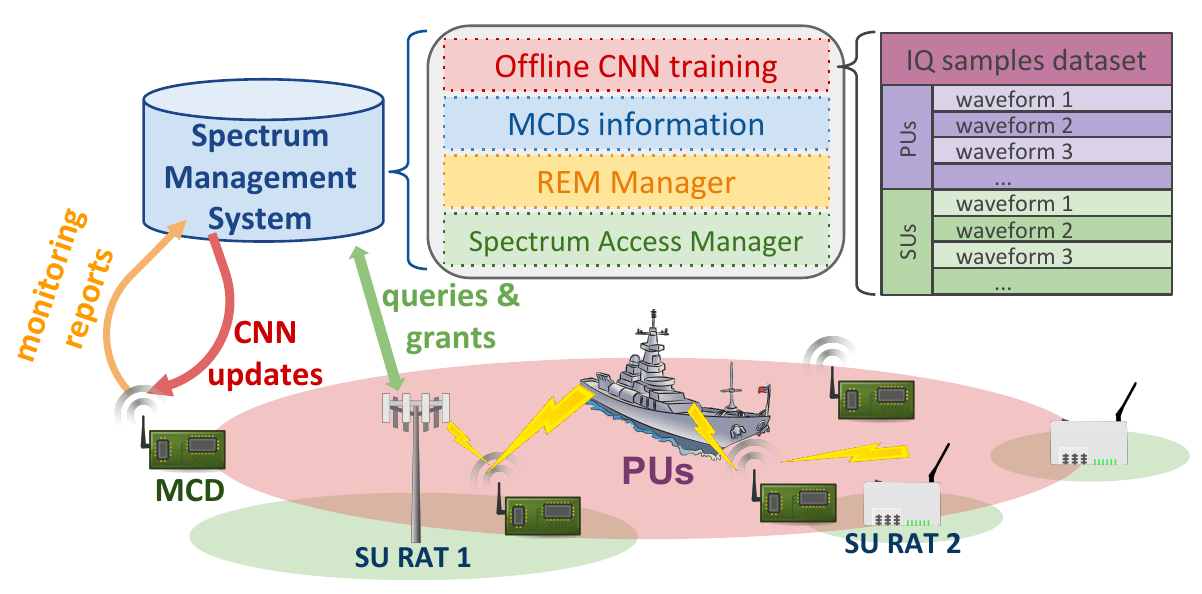}
  \caption{Illustration of the proposed deep learning framework for spectrum monitoring. \acp{mcd} report sensing results using \ac{cnn} models trained by the spectrum manager.\label{fig:scenario_paper}}
\vspace{-0.25in}
\end{figure}

One technical challenge for spectrum monitoring networks is in the selection of  sensing algorithms that enable \acp{mcd} to infer spectrum occupancy. Several of the existing works on \ac{rem} assume the use of received signal strength indicator (RSSI) or energy detection-based methods to delimit incumbents' protection zones\cite{gavrilovska2014enabling,ying2015incentivizing}. This may not, however, be an effective approach when multiple 
\ac{rats}, with different priorities for spectrum access and levels of protection from interference coexist in the same band. In such scenarios, there is a requirement for \acp{mcd} to be able to discriminate different users and technologies, which can only be achieved with more advanced signal classification algorithms.

Over the last decades, expert feature search and cyclostationary detection algorithms have dominated cognitive radio approaches on signal detection and classification \cite{swami2000hierarchical,kim2007cyclostationary,paisana2012alternative,ramkumar2009automatic}. However, the design of these specialized solutions has been proven to be time-demanding and inefficient, especially in 
license-exempt bands and some military-controlled bands, where a large number of signal types may coexist. With recent breakthroughs in machine learning, more flexible approaches based on neural networks started outperforming expert systems at specific tasks \cite{o2016convolutional}. However, these solutions have been primarily focused on the specific problem of modulation recognition (ModRec), which only represents a sub-task of the emitter classification problem in spectrum monitoring. In contrast to ModRec, in this paper, we consider a sharing scenario where RATs can employ adaptive modulation schemes, the incumbents and \acp{su} may follow incompatible channelization strategies, and overlapping between their signals might occur. One such scenario is spectrum sharing in radar bands, which is being considered by regulators and the wireless communication industry\cite{ITU-R2011,FCC2012,Drocella2015}.

In this paper, we assess the applicability of \acp{cnn} for spectrum monitoring, in particular, in the identification of radar signals. Our proposed framework, illustrated in Figure \ref{fig:scenario_paper}, consists of a spectrum management system that handles: (i) the channel access authorization of \acp{su}, (ii) \ac{rem} generation based on \acp{mcd}' sensing reports, (iii) and training of the \ac{cnn} models that the \acp{mcd} will employ to identify incumbent signals. This \ac{cnn} training procedure takes place offline, using datasets with all possible signal types, including incumbents and \acp{su}, that may access the spectrum. One advantage of our proposed framework is being software-based. In particular, the \acp{mcd}' \ac{cnn} models can be updated whenever new RATs are introduced in the targeted frequency band. Another advantage is obfuscation, as \acp{mcd} do not need explicit information about the radar systems' waveforms, which meets the secrecy requirements of some military bands.

We considered a radio environment where three types of technology may coexist:  radars as incumbents, and \ac{wlan} and commercial downlink \ac{lte} as \acp{su}. The main goal of this work is to build a machine learning-based signal classification model to successfully identify the presence of radar signals. To do this, we started by collecting and labeling a large dataset of RF measurements. We then assess the suitability of different signal representations (e.g. spectrograms and amplitude variation) to perform our classification task. Based on this analysis, we propose a unique signal representation that leverages amplitude and phase shift properties of radar signals. This representation allows our proposed \ac{cnn} model to achieve robust and reliable classification, as will be shown in the results section.

\newcommand{\TOA}{t^\textup{TOA}}
\newcommand{\TOAref}{t^\textup{TOA,ref}}
\newcommand{\PW}{\Delta t^\textup{pw}}
\newcommand{\PF}{f^c}
\newcommand{\PRI}{\Delta t^\textup{pri}}
\newcommand{\PA}{A}
\newcommand{\phasediff}{\Delta\phi}

\section{Spectrum Monitoring in Radar Systems}

In this section, we discuss the main challenges associated to the identification of radar signals, in low SNRs or when overlapped with other communication signals. In particular, we briefly characterize different features of radar pulses, and what makes them distinguishable in comparison to signals of current commercial broadband \ac{rats} such as \ac{wlan} and \ac{lte}. This effort will provide some intuition about what data representation should be utilized by \ac{cnn} models.



\subsection{Radar Signal Model}

\begin{figure*}
\centering
\includegraphics[width=0.8\textwidth]{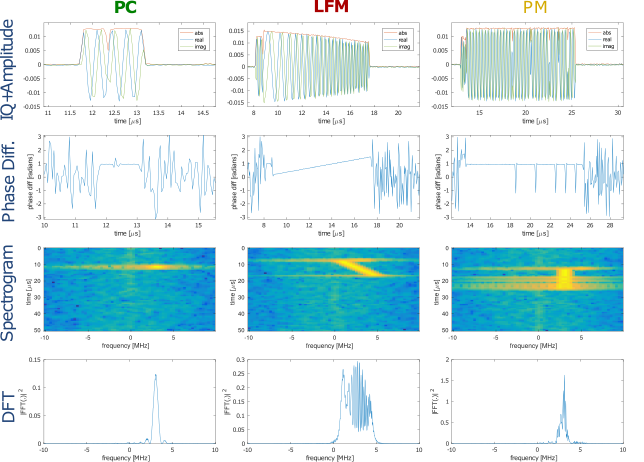}
\caption{Examples of possible radar pulse waveforms.\label{fig:pulse_waveforms}}
\vspace{-0.3in}
\end{figure*}

To construct our system model, we imagine a single \ac{mcd} at a fixed position, receiving signals from several communication systems in its surroundings. A radar system is suddenly turned on and it starts transmitting pulses through a directive antenna. To create an \ac{rem} of the radar system's coverage, the \ac{mcd} has to be able to detect its pulses when they are at their highest amplitude. This generally corresponds to the instants of time when the radar antenna main beam directly illuminates the \ac{mcd}.

The samples received by the \ac{mcd} from a single radar in an ideal channel can be modeled as follows,
\begin{equation}
x[n] = \sum_{m=-\infty }^{+\infty}\PA_{m}.\left(p[n-m\PRI_{m}] \ast h_m[n]\right) .e^{2\pi jn\PF_{m}/f_0}.
\end{equation}
Here, $n$ is the sample index and $m$ is the pulse index. The pulse shape $p[\cdot]$ 
defines the \ac{ipm} and \ac{pw}. The pulse amplitude ($\PA$) suffers sharp variations with time due to the radar antenna steering. The pulse centre frequency $\PF$ causes a rotating phase shift in the received IQ samples. The \ac{pri} ($\PRI$) of the radar defines the interval between pulses' \ac{toa}. The radio channel $h_m[\cdot]$ has the effect of dispersing the pulse shape $p[\cdot]$ over time.

\subsection{Radar Signal Representations}
In Figure \ref{fig:pulse_waveforms}, we illustrate three examples of commonly used radar \acp{ipm}: \ac{pc}, \ac{lfm} and Barker BPSK \ac{pm}. These samples were generated through the radar waveform emulator described in \cite{santos2015context} and transmitted over the air. The spectrum analyzer was operating at a sampling rate of 20 MS/s and misaligned with the transmitter by 3 MHz. The first row of Figure \ref{fig:pulse_waveforms} shows the IQ and amplitude of different radar pulses over time. The following rows illustrate the phase difference ($\phasediff$) between consecutive IQ samples, spectrogram, and magnitude square of the signal's DFT, respectively. 

The radar pulse waveforms are quite recognizable under any of the chosen representations. However, each representation provides a different level of robustness to non-ideal channel and radio front-end effects. Taking for instance Amplitude+IQ subfigures, we can observe that the amplitude envelope of the received pulses is not perfectly constant over time. In the PM phase difference plot, it can also be seen that the phase shift of the BPSK pulse does not ever reach 180 degrees. These phenomena are generally a consequence of multipath or the limited receiver's sampling rate that filters out high frequency components of the pulses. In \cite{Paisana2017}, the authors show that in long distance NLOS outdoor settings, radar pulses can incur even further amplitude and IQ distortions than the ones illustrated in Figure \ref{fig:pulse_waveforms}.

Misalignments between the transmitter and receiver's centre frequencies, on the other hand, have the effect of offsetting the pulse position in the DFT and spectrogram plots, and altering the mean of its $\phasediff[n]$ curve. This mean value can be computed as follows,
\vspace{-0.2in}

\begin{align}
\begin{split}
\Delta\phi=&
\measuredangle\left(p[n].e^{2\pi jn\frac{\PF}{f_0}}.\left(p[n-1].e^{2\pi j(n-1)\frac{\PF}{f_0}}\right)^*\right)\\
=&\measuredangle \left(p[n].p^*[n-1].e^{-2\pi j\frac{\PF}{f_0}}\right)\\
=&-2\pi\frac{\PF}{f_0}\measuredangle \left(p[n].p^*[n-1]\right).
\end{split}
\end{align}

\vspace{-0.07in}
\noindent where $\measuredangle$ is the angle operator, and $\measuredangle \left(p[n].p^*[n-1]\right)$ is the $\phasediff[n]$ when the transmitter and receivers are exactly frequency aligned. 
In IQ versus time representations, on the other hand, any frequency misalignment will lead to a significant alteration in the pulse curve shape. 

Classification of radar signals based on DFT may provide very poor performances at low SNRs, due to the generally very short duration of radar pulses compared to the DFT window. In addition, frequency selective multipath or the presence of multiple pulses per DFT window will significantly affect the shape of the resulting DFT.

\subsection{SU Signal Representations}
\begin{figure}
\centering
\includegraphics[width=0.9\columnwidth]{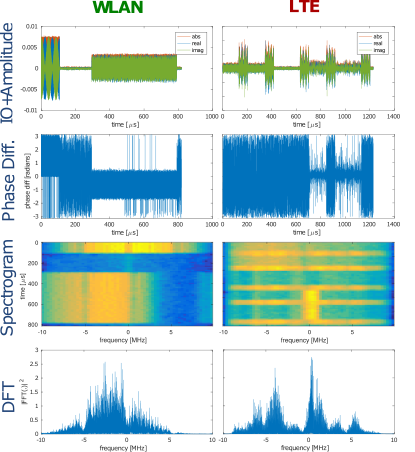}
\caption{Illustrations of WLAN and LTE waveforms.\label{fig:ltewifi_waveforms}}
\vspace{-0.2in}
\end{figure}

We selected two types of commercial technologies as secondary users: \ac{wlan} and DL-\ac{lte}. In Figure \ref{fig:ltewifi_waveforms}, we show representative examples of the IQ and amplitude over time, phase difference, spectrogram, and DFTs of these two signals, captured from the commercial bands of 2.462 GHz and 906 MHz, respectively. The spectrum analyzer's sampling rate is set to 20 MS/s. As illustrated in the \ac{wlan} spectrogram, multiple users can contend for access to \ac{wlan} bands, with different frequency alignments, and transmissions are expected to occur in a bursty fashion using the CSMA/CA protocol. Although licensed \ac{lte} eNB are generally deployed more sparsely and transmit in a continuous manner, their transmit power allocation over time and frequency varies according to the load and scheduler of the cell. The \ac{lte} bursty wideband emissions observed in Figure \ref{fig:ltewifi_waveforms} are caused by the \ac{lte} reference signaling, and becomes more noticeable the less loaded the cell is. \ac{lte} systems also have configurable bandwidth, ranging from 1.4 MHz to 20 MHz without carrier aggregation.

The transmission burstiness and the superposition of misaligned \ac{wlan} signals can unpredictably alter the shape of the DFTs, making it an ill-suited representation for \ac{wlan} signal classification. On the other hand, both \ac{wlan} phase difference and the spectrogram plots show quite distinctive features compared to radar pulses. The \ac{wlan} signal amplitude is fairly constant over time and its packets generally display longer durations than radar pulses. 

The flexible scheduling schemes of \ac{lte} systems lead to an almost infinite number of possible DFT shapes, spectrogram, phase difference, and IQ representations. The short durations of its OFDM symbols may also cause non-constant amplitude bursts of around 66$\mu\textup{s}$ of duration, which can be challenging to discriminate from radar pulses, when employing a hand-coded energy-based pulse detection approach. The most distinctive features of \ac{lte}'s bursts of energy, compared to radar pulses, are their wider instantaneous phase and frequency spreads.

\section{CNN for radar signal classification}
\label{sec:deep_learning}

\subsection{Data Collection}

Our data collection setup consisted of two USRP N210 front-ends. One USRP, which we denote as SDR-Radar, is used for generating and transmitting radar pulses every 1 ms. The second USRP, called SDR-MCD, operates as an \ac{mcd}, receiving and storing samples into a file at 20 MS/s. We performed this sample collection over three different bands: \ac{lte} 906 MHz, ISM 2462 MHz, and the 2300 MHz band, where the latter represented a radio medium that is free of interference. To increase our \ac{cnn} model's robustness to channel and radio front-end effects and increase its capability to generalize to multiple radar waveforms, we altered the parameters of the SDR-Radar in a random fashion. Table \ref{table:collected_data} shows the set of parameter configurations used to collect radar signals over different bands. These parameters include the \ac{ipm}, \ac{pw}, frequency excursion $f^e$, and frequency misalignment $\Delta f$ between the SDR-Radar and SDR-MCD.

We partitioned each collected file into chunks of 1024 IQ samples. Table \ref{table:spectrogramdataset} shows the total number of chunks used during the training and testing phases. Here, class 0 includes chunks with radar-only, radar+\ac{wlan}, and radar+\ac{lte} samples. Class 1 includes \ac{lte}-only, \ac{wlan}-only, and noise. All collected chunks are checked manually and mislabelled data is removed.

\begin{figure*}[h]
\centering
  \includegraphics[scale=0.6]{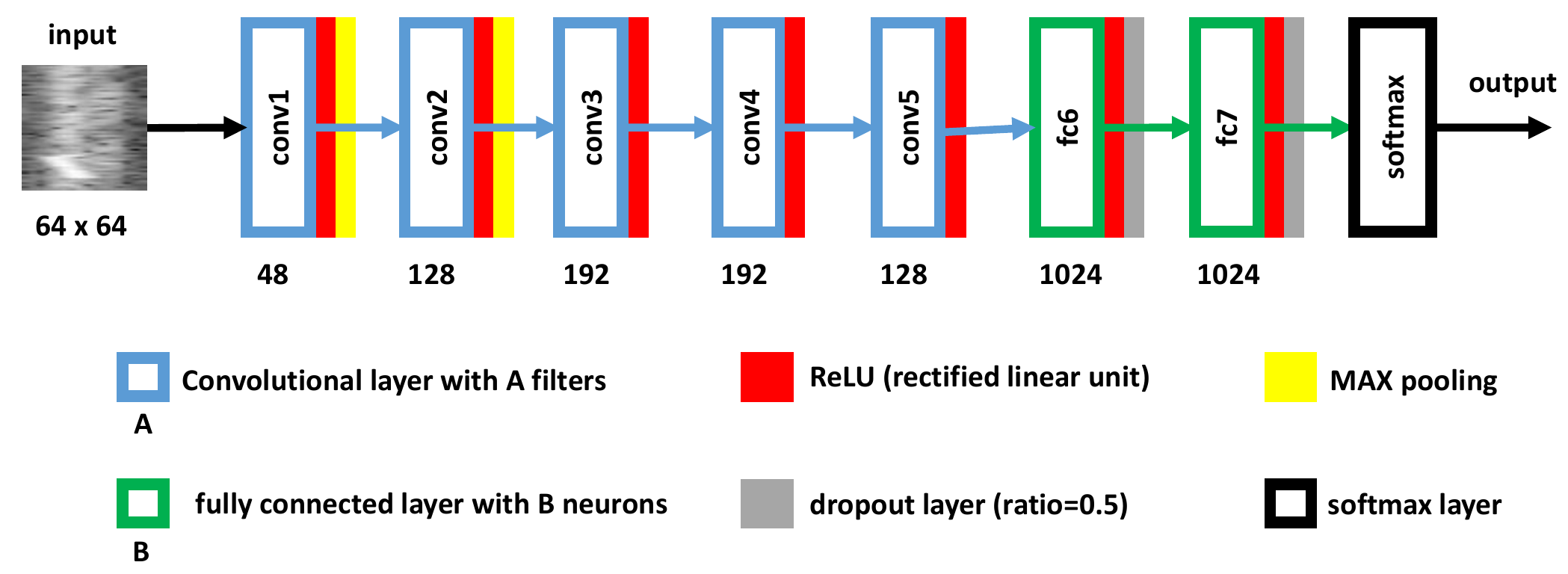}
\vspace{-0.07in}
  \caption{Proposed CNN architecture.\label{fig:CNN}}
\vspace{-0.4in}
\end{figure*}

\begin{table}[]
\centering
\caption{Set of radar parameters and bands utilized during the \ac{cnn} training and testing phases.}
\label{table:collected_data}
\begin{tabular}{|c|c|c|c|c|}
\hline
\textbf{Parameters}  & \multicolumn{4}{c|}{\textbf{Values}} \\ \hline
$\Delta f$ {[}MHz{]} & \multicolumn{4}{c|}{\{-6,-3,0,3,6\}} \\ \hline
IPM                  & PC   & PC   & LFM  & PM (Barker 13)  \\ \hline
PW {[}$\mu$s{]}      & 2    & 10   & 10   & 10              \\ \hline
$f^e$ {[}MHz{]}      & 0    & 0    & 4    & 0               \\ \hline
\end{tabular}
\end{table}

\begin{table}
\centering
\caption{Number of 1024-sized IQ chunks used per scenario\label{table:spectrogramdataset}}
\begin{tabular}{ |c|c|c|c| } 
\hline
class & training dataset & testing dataset \\
\hline
0: radar signal present & 20000 & 4000 \\
\hline
1: radar signal absent & 20000 & 4000 \\
\hline
\hline
Total & 40000 & 8000 \\
\hline
\end{tabular}
\vspace{-0.2in}
\end{table}

\subsection{CNN Models: Design and Performance Evaluation}

Inspired by the aforementioned discussion, as well as the visual characteristics of the radar pulses in Figure \ref{fig:pulse_waveforms} and \ac{wlan} and \ac{lte} signals in Figure \ref{fig:ltewifi_waveforms}, we infer that spectrograms, amplitude and phase difference representations are the most suitable for our \ac{cnn} solution. These representations are not sensitive to frequency misalignments and phase shifts. In addition to this, radar signals have unique characteristics under these transformations, which helps \ac{cnn} models to perform reliable classification.

For simplicity, we will start with spectrograms and show how they can be used as data representations. The same framework can be extended to other representations in a straightforward manner. Figure \ref{fig:CNN} shows our proposed CNN architecture. Here, the input of the network is a spectrogram of size $64\times64$. To reach the output of the network, the spectrogram is fed to five consecutive convolutional layers, followed by 2 fully connected layers. The $i$th convolutional layer consists of $N_i$ filters with dimension of $11 \times 11$, $5 \times 5$, $3\times3$, $3\times3$, $ 3 \times 3$ for $i=1,2,..,5$, respectively. The output of the last fully connected layer is fed to a softmax classifier to compute the probability $P(y=k|x;\theta)$ for $k \in \{0,1\}$, where $x$ denotes the input spectrogram, and $\theta$ denotes model parameters. $k=0$ denotes the presence of radar signals.
The CNN model is trained on 40K spectrograms, for 25K iterations with batch size of 50. The training is performed using the stochastic gradient descent algorithm in which the weights of the network are updated after each iteration to minimize classification errors. The learning rate starts with 0.01 and is divided by a factor of 10 every 5K iterations. The momentum is set to 0.9 and the regularization parameter is set to 0.001 to avoid over-fitting. The model is trained using the Caffe framework \cite{caffe} on a powerful GPU (Tesla K40c). After training, the model is tested over 8K spectrograms (test dataset) achieving classification accuracy of $98.6\%$.

The same CNN model can be used for other data representations with slight modifications to reflect the change in the input size. It is possible to build a CNN model for the amplitude representation (A-CNN), and another model for the phase difference $\Delta \phi$ (P-CNN). We found that both models achieve worse performance than the spectrogram-based CNN model (S-CNN). However, we also observe that amplitude and phase difference representations are correlated in radar signals (see Fig. \ref{fig:ltewifi_waveforms}- first two rows). We therefore combine both representations and feed them as input to our CNN model. The intuition behind this is to allow the CNN model to capture the inter-dependencies between phase difference and amplitude signals present in radar pulses that the A-CNN and P-CNN models were not capable of.
To achieve this, first, we normalize both representations to have values from 0 to 1. Amplitude representations are divided by their maximum values, while phase difference representations are divided by $2 \pi$. We then concatenate the two representations in a $1024\times2$ matrix. For convenience, we reshape this matrix into $64\times64\times2$. Now, the same CNN architecture can be used, however, with a slight variation in the first convolutional layer (filter dimensions are set to $2 \times 11 \times 11$). All other network layers and training parameters remain the same. The trained model achieved classification accuracy of $99.6\% $ on our testing dataset. Although both models can provide high classification accuracies, we observe that the Amplitude+Phase difference-based CNN model (AP-CNN) achieves better performance than the S-CNN model. This is due to the fact that AP-CNN model, unlike S-CNN, utilizes phase information. 


\begin{figure}[th]
\centering
  \includegraphics[scale=0.6]{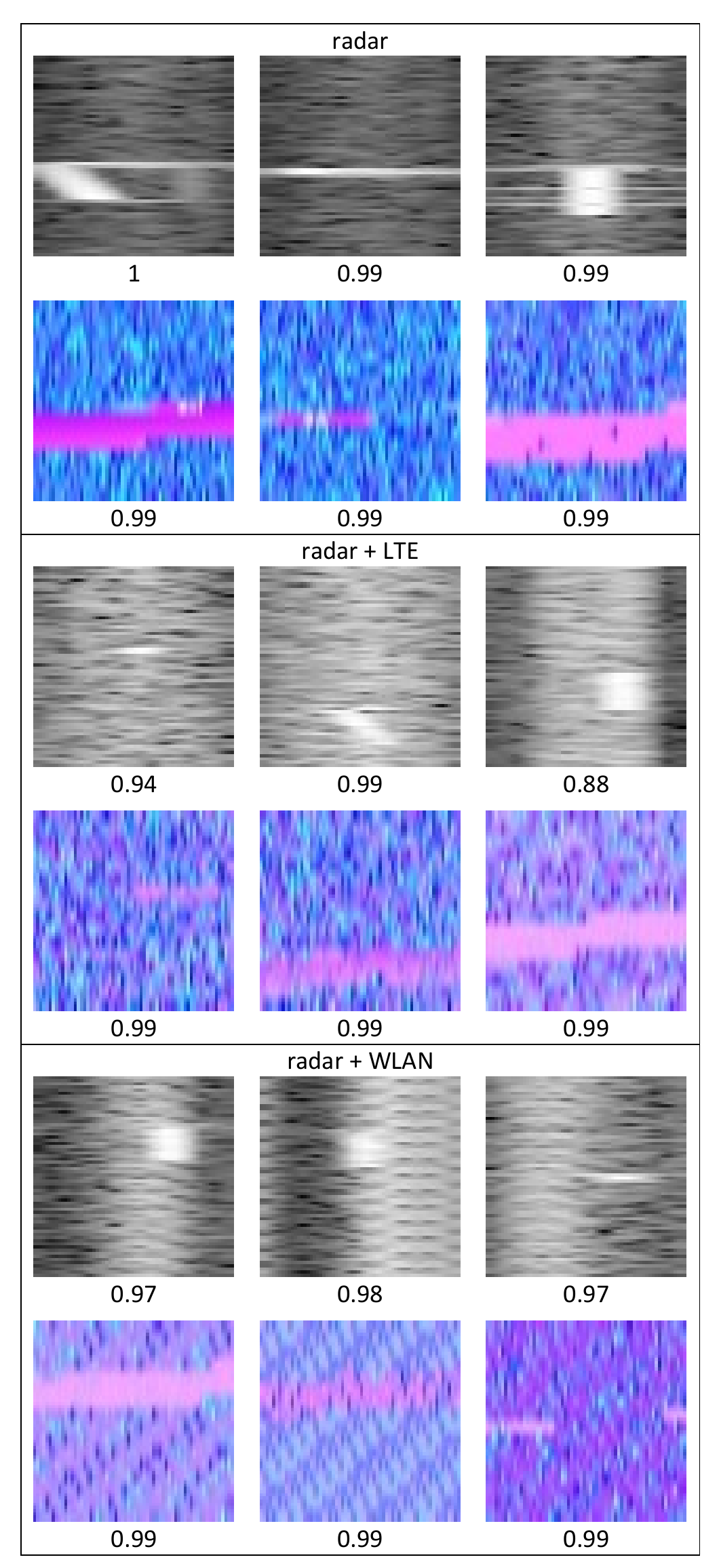}
  \caption{Examples from our testing dataset when radar signals are present. $P_S$ and $P_{AP}$ are shown below spectrograms and AP representations, respectively.\label{fig:R}}
\vspace{-0.25in}
\end{figure}

\begin{figure}[th]
\centering
  \includegraphics[scale=0.6]{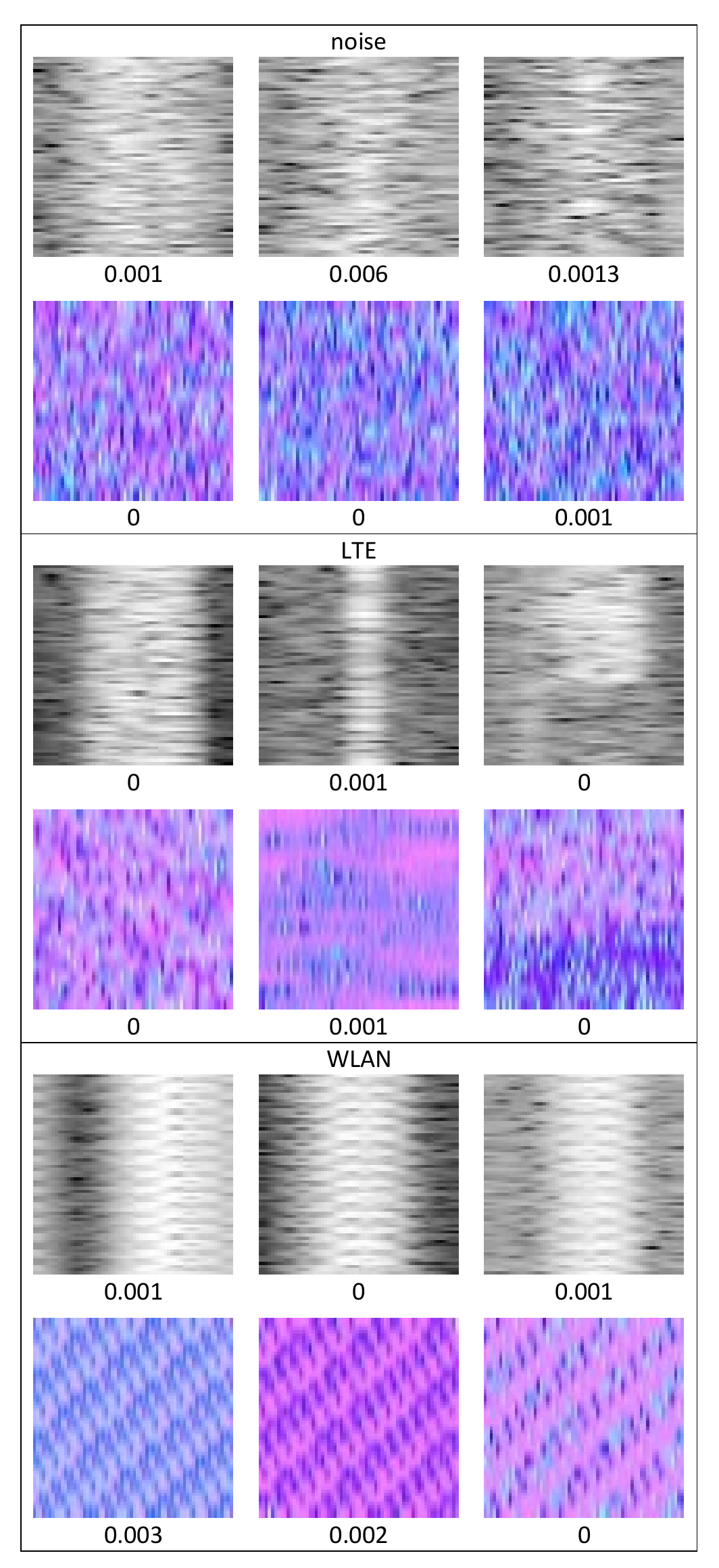}
  \caption{Examples from our testing dataset when radar signals are absent. $P_S$ and $P_{AP}$ are shown below its associated spectrograms and AP representations, respectively.\label{fig:Rbar}}
\vspace{-0.25in}
\end{figure}

\section{Experimental results}
\label{sec:experimental_results}

\subsection{Discrimination between Radar Signals and SUs}

Let $P_{S}$, $P_{AP}$ denote the probability $P(y=0|x;\theta_S)$ and $P(y=0|x;\theta_{AP})$, where $\theta_S$, $\theta_{AP}$ denote model parameters of the S-CNN and the AP-CNN models, respectively. Figures \ref{fig:R} and \ref{fig:Rbar} show few examples from our testing dataset and the generated CNN results. The first row of each section represents 3 spectrograms, while their corresponding AP-representations are shown in the second row. The numbers below the spectrograms and AP-representations indicate the obtained $P_{S}$ and $P_{AP}$ values, respectively. Here, for visualization purposes, we show AP-representations as RGB images. Normalized amplitude is assigned to the red channel, normalized phase difference is assigned to the green channel and the blue channel is set to 1. We observe that both CNN models successfully identified the presence/absence of radar signals in the tested cases.

\subsection{Sensitivity to Noise}

\begin{figure}
\centering
\includegraphics[width=0.85\columnwidth]{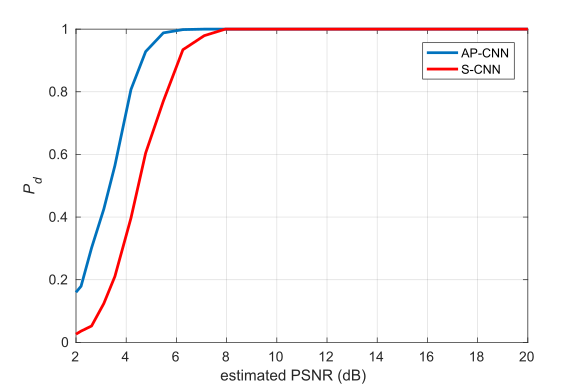}
\caption{Probability of pulse detection for the \ac{lfm} waveform at different PSNR levels.\label{fig:LFM}}
\vspace{-0.2in}
\end{figure}

\begin{figure}
\centering
\includegraphics[width=0.85\columnwidth]{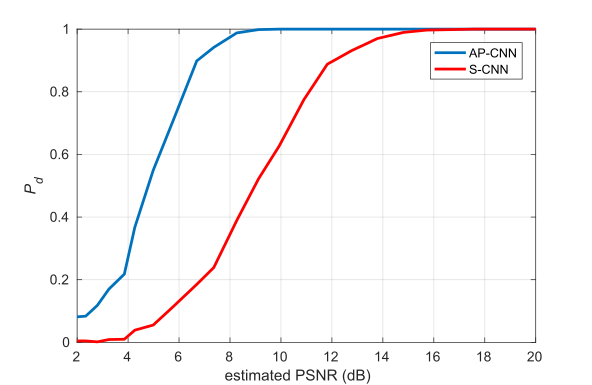}
\caption{Probability of pulse detection for the \ac{pc} waveform with a duration of 2 $\mu$s at different PSNR levels.\label{fig:PC1}}
\vspace{-0.2in}
\end{figure}

\begin{figure}
\centering
\includegraphics[width=0.85\columnwidth]{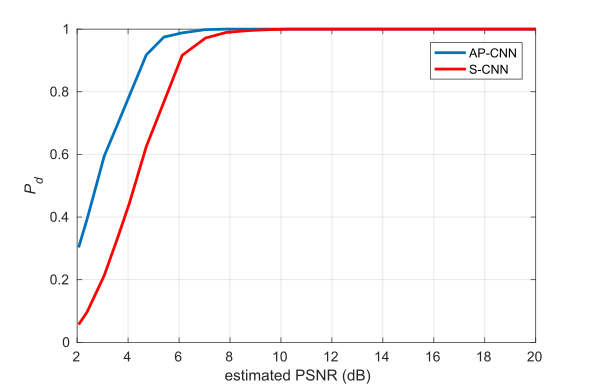}
\caption{Probability of pulse detection for the \ac{pc} waveform with a duration of 10 $\mu$s at different PSNR levels.\label{fig:PC2}}
\vspace{-0.2in}
\end{figure}

\begin{figure}
\centering
\includegraphics[width=0.85\columnwidth]{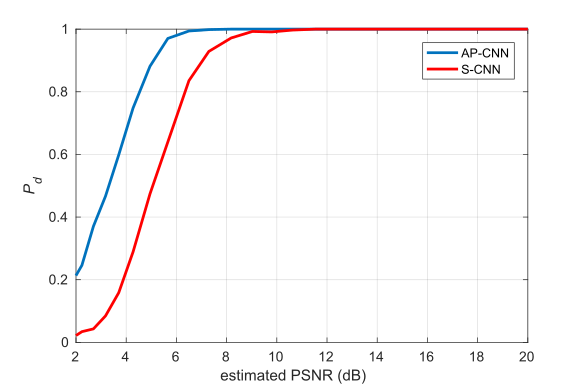}
\caption{Probability of pulse detection for the \ac{pm} waveform at different PSNR levels.\label{fig:PM1}}
\vspace{-0.2in}
\end{figure}

In this section, we test the detection performance of both S-CNN and AP-CNN models under different noise levels. For this test, our dataset consists of 18 sets, each containing 670 IQ chunks of radar+noise samples. Each set was collected using different SDR-Radar USRP N210 transmit gains. The transmitted waveforms were the same as the ones shown in Table \ref{table:collected_data}. To estimate the average peak SNR (PSNR) of each set of chunks, we computed the mean power of the radar pulses' samples and the mean power of the remaining samples within each set. $P_d$ is the detection probability of radar pulses. 

Figures \ref{fig:LFM}, \ref{fig:PC1}, \ref{fig:PC2} and \ref{fig:PM1} show the obtained results using S-CNN and AP-CNN models. It is clear that the AP-CNN outperforms the S-CNN model, especially at low PSNR values, and for short pulse durations. The main reason behind this success is the fact that the phase difference information embedded in radar pulses helps the \ac{cnn} model to distinguish them from the general impulsiveness of thermal noise.

\section{Conclusion and Future Work}
In this work, we presented a spectrum monitoring framework for radar bands. Our goal was to detect the presence of radar signals even in the case of simultaneous transmission of \ac{lte} and \ac{wlan} systems. This is achieved through deep \acp{cnn}. We tested the performance for different data representations and concluded that our proposed Amplitude+Phase Difference representation enables \acp{cnn} models to obtain high classification accuracy and it is more robust to noise.

As future work, we intend to extend our dataset to include other \ac{rats}, such as LAA-LTE, Multefire, and NB-IoT as \acp{su}, and more radar waveforms. Our training and testing datasets will be available online for benchmarking and further research. Another direction for this work is to utilize our new trained CNN models in wireless systems' control and decision making such as the work presented in \cite{Paisana2017challenge}.

\label{sec:conclusion}


\section*{Acknowledgment}
This publication has emanated from the research supported by the European Commission Horizon 2020 Program under grant agreements no. 688116 (eWINE) and no. 732174 (ORCA), and co-funded under the European Regional Development Fund from Science Foundation Ireland under Grant Number 13/RC/2077 (CONNECT). Deep learning models  were trained on the machines maintained by the Trinity Center for High Performance Computing.

\bibliographystyle{IEEEtran}
\bibliography{bibliography} 

\end{document}